\documentstyle[aps,preprint]{revtex}

\begin{document}
\draft
\author{R.Zh.~ SHAISULTANOV\thanks{%
Email:shaisultanov@inp.nsk.su}}
\address{Budker Institute of Nuclear Physics \\
630090, Novosibirsk 90, Russia}
\title{Backreaction in scalar QED, Langevin equation and Decoherence
functional}
\date{   }
\maketitle

\begin{abstract}
Using the Schwinger-Keldysh (closed time path or CTP) and Feynman-Vernon
influence functional formalisms we obtain a Langevin equation for the
description of the charged particle creation in electric field and of
backreaction of charged quantum fields and their fluctuations on time
evolution of this electric field. We obtain an expression for the influence
functional in terms of Bogoliubov coefficients for the case of quantum
electrodynamics with spin 0 charged particles. Then we derive a CTP
effective action in semiclassical approximation and its cumulant expansion.
An intimate connection between CTP effective action and decoherence
functional will allow us to analyze how macroscopic electromagnetic fields
are ``measured'' through interaction with charges and thereby rendered
classical.
\end{abstract}

\pacs{PACS numbers 03.65.Db, 03.70.+k, 05.40.+j, 11.15.Kc}

\newpage

\section{Introduction}

Nonequilibrium aspects of quantum field theory are beginning to receive
considerable attention in recent years. Among them are dissipation and
decoherence in quantum cosmology \cite{di}, structure formation in
inflationary cosmology \cite{s1,s2,s3}and time evolution of a nonequilibrium
chiral phase transition \cite{s4} to name only a few.

An interesting problem of similar nature is a problem of backreaction of
particle production in a dynamical background field. Backreaction of quantum
processes like particle creation in cosmological spacetimes \cite{Z} has
been considered by many authors to understand how quantum effects affect the
structure and dynamics of the universe near Planck time\cite{di1}.
E.Calzetta and B.L.Hu \cite{di2} used Schwinger-Keldysh (or
closed-time-path, CTP ) functional formalism to derive a real and causal
equation of motion (the semiclassical Einstein equation) describing the
backreaction of particle production on classical effective geometry. In this
equation one can identify a nonlocal kernel in the dissipative term whose
integrated dissipative power has been shown to be equal to the energy
density of the total number of particles created, thus establishing the
dissipative nature of the backreaction process. Then in \cite{q} Hu pointed
out that a Langevin-type equation is what should be expected, and predicted
that for quantum fields a colored noise source should appear in the driving
term. Calzetta and Hu \cite{q1} show how noise and fluctuations can be
attained with the CTP formalism together with dissipation and decoherence.
Hu and Matacz \cite{q2} used cumulant expansion of the influence functional
of Feynman and Vernon ,derived in terms of the Bogoliubov coefficients, to
extract the noise associated with the matter field and to derive an
Einstein-Langevin equation. In the Feynman-Vernon formalism \cite{fe,fe1,fe2}
the effects of noise and dissipation can be extracted from the imaginary and
real parts of the influence functional. Also it was shown \cite{q,hs} that
backreaction effect on the dynamics of space-time can be viewed as a
manifestation of fluctuation-dissipation relation . Thus Hu and
collaborators have extended the old framework of semiclassical gravity,
based on Einstein equation with expectation value of energy-momentum tensor
as a source , to that based on a Langevin equation which describes also the
fluctuations of matter fields and spacetime ( see recent review \cite{hu} ).

In this article we will study the quantum non-equilibrium effects of pair
creation in strong electric fields. Backreaction of pair creation on
electric field was recently discussed by Cooper, Mottola at all \cite{klu}.
They derived the semiclassical Maxwell equation, carry out its
renormalization and numerically solve it for some initial conditions in 1+1
dimensions.Their numerical results clearly exhibits the decay of
 the electric field because of screening by the produced particles.
 We wish to make a step further and derive a Langevin equation,
taking into account a noise from quantum matter fields. To this purpose we
will use some mixture of Schwinger-Keldysh (CTP) and Feynman-Vernon
influence functional formalisms. It is important to note here that
phenomenological equations of motion with noise term can also be derived
using decoherence functional formulation of quantum mechanics. This was done
for some model quantum systems in \cite{gh,brun}

This paper is organized as follows: In Sec.\ref{f2} we give a brief review of
 CTP
functional formalism , mainly to introduce notations. All essential details
can be found in \cite{hu,w,we,wee}. In Sec.\ref{f3} we will obtain an
 expression
for the influence functional in terms of Bogoliubov coefficients for the
case of quantum electrodynamics with spin 0 charged particles. Then in
 Sec.\ref{f4}
we will obtain a CTP effective action in semiclassical approximation and its
cumulant expansion. It is the main result of this paper. We will apply it
for study of two interesting problems. First this result will allow us to
analyze the backreaction of created charged particles on electric field and
to derive a Langevin equation, which take into account a noise from quantum
matter fields. Then in Sec.\ref{f5} we will use an intimate connection between
CTP\
effective action and decoherence functional \cite{q1,hu} to analyze how
macroscopic electromagnetic fields are ``measured'' through interaction with
charges and thereby rendered classical.

\section{ The Closed Time Path Functional Formalism in Quantum Field Theory }
\label{f2}

Usually in quantum field theory our interest is in obtaining the amplitudes
of transition from in-states to the out-states. But in many cases, mainly in
statistical physics, we are concerned with expectation values of physical
quantities at finite time. To solve such initial value problems Schwinger
has invented close time path (CTP) formalism.

Let us consider the expectation value of an arbitrary operator $A$:

\begin{equation}
\label{eq:wy1}<A>\left( t\right) =Tr\ \rho \left( t\right) \ A
\end{equation}

Here $\rho $ is the density matrix that describes the ( mixed ) state of the
system. The density matrix does not necessarily have to commute with the
Hamiltonian, in which case it describes a non-equilibrium state. Using
relation\quad $\rho \left( t\right) =U\left( t,0\right) \,\rho \left(
0\right) \,U^{-1}\left( t.0\right) $ where $U\left( t,0\right) $ is the
evolution operator, inserting the identity operator $1=U\left( t,T\right) $ $%
U\left( T,t\right) $ we obtain

\begin{equation}
\label{eq:qw}
\begin{array}{c}
<A>\left( t\right) =Tr
\text{ }\rho \left( 0\right) \text{ }U^{-1}\left( t,0\right) \text{ }A\text{
}U\left( t,0\right) = \\ Tr\text{ }\rho \left( 0\right) \text{ }U\left(
0,T\right) \text{ }U\left( T,t\right) \text{ }A\text{ }U\left( t,0\right)
\end{array}
\end{equation}
\

Equation (\ref{eq:qw}) can be pictured as describing the evolution of the
system from $0$ to $t$, inserting the operator $A$ , evolving further to
some large time $T$ \ (in practice, $T\rightarrow \infty $), and then
backwards from $T$ \ to $0$. The insertion of operator may be achieved by
introducing external sources coupled to the particular operator. This
suggests the definition of the CTP generating functional

\begin{equation}
\label{eq:ww}Z\left[ J_{+,}J_{-}\right] \equiv \exp iW\left[
J_{+},J_{-}\right] =Tr\ \rho \left( 0\right) U\left( 0,T,J_{-}\right)
U\left( T,0,J_{+}\right)
\end{equation}

In the path integral representation we have

\begin{equation}
\label{eq:wy2}Z\left[ J_{+},J_{-}\right] =\int D\phi _1D\phi _2D\phi \text{ }%
\left\langle \phi _1\left| \rho \right| \phi _2\right\rangle
\,\int\limits_{\phi _1}^\phi D\phi _{+}\int\limits_{\phi _2}^\phi D\phi
_{-}\exp i\int\limits_0^Tdt\left\{ L\left[ \phi _{+}\right] -L\left[ \phi
_{-}\right] +J_{+}\phi _{+}-J_{-}\phi _{-}\right\}
\end{equation}

The expectation values can be obtained as
\begin{equation}
\label{eq:wy3}\bar \phi _{+}=\frac{\delta W}{\delta J_{+}}%
\,\,\,\,\,\,\,,\,\,\,\bar \phi _{-}=\frac{\delta W}{\delta J_{-}}
\end{equation}

Then the CTP effective action is

\begin{equation}
\label{eq:wy4}\Gamma _{CTP}\left[ \bar \phi _{+},\,\bar \phi _{-}\right]
=W\left[ J_{+},J_{-}\right] -J_{+}\bar \phi _{+}+J_{-}\,\,\bar \phi _{-}
\end{equation}

The equations of motion are
\begin{equation}
\label{eq:wy5}\frac{\delta \Gamma _{CTP}}{\delta \bar \phi _{+}}%
=-J_{+}\,\,\,,\,\,\,\,\frac{\delta \Gamma _{CTP}}{\delta \bar \phi _{-}\,}%
=J_{-}
\end{equation}

The physical situations correspond to solutions of the homogeneous equations
at$\bar \phi _{+}=\,\bar \phi _{-}$ . Then equations are real and causal.

To apply this formalism to our situation we should substitute the $\phi $
field by the pair $\psi $ and $\sigma .$ We will be interested in
expectation values of $\,\psi $ only, so we do not couple the $\sigma $
field to an external source. Also we assume that the initial density matrix
factorizes $\rho =\rho _\psi \,\,\rho _\sigma $ . Then we have
\begin{equation}
\label{eq:zzz}
\begin{array}{c}
Z\left[ J_{+},J_{-}\right] =\int D\psi _1D\sigma _1D\psi _2D\sigma
_2\,\left\langle \psi _1\left| \rho _\psi \right| \psi _2\right\rangle
\,\,\left\langle \sigma _1\left| \rho _\sigma \right| \sigma _2\right\rangle
\, \\
\int D\psi D\sigma \,\int\limits_{\psi _1}^\psi D\psi
_{+}\,\int\limits_{\sigma _1^{}}^\sigma D\sigma _{+}\int\limits_{\psi
_2}^\psi D\psi _{-}\,\int\limits_{\sigma _2^{}}^\sigma D\sigma _{-}\,\exp
\,i\,\int\limits_0^T\,dt\,\left\{ {}\right. L_\psi \left[ \psi _{+}\right]
-L_\psi \left[ \psi _{-}\right] +J_{+}\,\psi _{+}-J_{-}\psi _{-}+ \\
L_\sigma \left[ \sigma _{+}\right] -L_\sigma \left[ \sigma _{-}\right]
+L_{int}\left[ \psi _{+}\,,\sigma _{+}\right] -L_{int}\left[ \psi
_{-}\,,\sigma _{-}\right] \left. {}\right\} = \\
\int D\psi _1D\psi _2\,\left\langle \psi _1\left| \rho _\psi \right| \psi
_2\right\rangle \,\int D\psi \,\int\limits_{\psi _1}^\psi D\psi
_{+}\,\int\limits_{\psi _2}^\psi D\psi _{-}\,\exp
\,i\,\int\limits_0^T\,dt\,\left\{ {}\right. L_\psi \left[ \psi _{+}\right]
-L_\psi \left[ \psi _{-}\right] + \\
J_{+}\,\psi _{+}-J_{-}\psi _{-}\left. {}\right\} \,\Phi \left[ \psi
_{+}\,,\psi _{-}\right]
\end{array}
\end{equation}

where $\Phi \left[ \psi _{+}\,,\psi _{-}\right] $ is the so called influence
functional
\begin{equation}
\label{eq:inf}
\begin{array}{c}
\Phi \left[ \psi _{+}\,,\psi _{-}\right] \equiv \exp \,i\,S_{IF}\left[ \psi
_{+}\,,\psi _{-}\right] =\int D\sigma _1D\sigma _2\,\,\left\langle \sigma
_1\left| \rho _\sigma \right| \sigma _2\right\rangle \int\limits_{\sigma
_1^{}}^\sigma D\sigma _{+}\,\int\limits_{\sigma _2^{}}^\sigma D\sigma _{-}\,
\\
\exp \,i\,\int\limits_0^T\,dt\left\{ L_\sigma \left[ \sigma _{+}\right]
-L_\sigma \left[ \sigma _{-}\right] +L_{int}\left[ \psi _{+}\,,\sigma
_{+}\right] -L_{int}\left[ \psi _{-}\,,\sigma _{-}\right] \right\} = \\
Tr\,\left[ U\left( T,0;\psi _{+}\right) \,\,\rho _\sigma \left( 0\right)
\,U^{-1}\left( T,0;\psi _{-}\right) \right]
\end{array}
\end{equation}

It is now easy to show ,using (\ref{eq:zzz}) and (\ref{eq:inf}) ,that in
semiclassical approximation CTP effective action has the form
\begin{equation}
\label{eq:eff}\Gamma _{CTP}\left[ \psi _{+}\,,\psi _{-}\right] =S\left[ \psi
_{+}\right] -S\left[ \psi _{-}\right] +S_{IF}\left[ \psi _{+}\,,\psi
_{-}\right]
\end{equation}

{}From this relation one may derive the semiclassical equations of motion for
the expectation values of the $\psi $ field. It is worth noting that
\begin{equation}
\label{eq:efff}\Gamma _{CTP}\left[ \psi _{+}\,,\psi _{-}\right] =-\Gamma
_{CTP}^{*}\left[ \psi _{-}\,,\psi _{+}\right] \,\,\text{ and }\,\,\,\Gamma
_{CTP}\left[ \psi \,,\psi \right] \equiv 0.
\end{equation}

\section{Influence functional for scalar QED} \label{f3}

In this section we wish to find the influence functional in terms of
Bogoliubov coefficients (as in \cite{q2} ). Since we are dealing here with
the case of scalar QED the pair of fields $A$ and $\varphi $ will play a
similar role as $\psi $ and $\sigma $ in Sec.2 respectively. The influence
functional have now the form
\begin{equation}
\label{eq:u1}\Phi \left[ \,A^{\prime },A\right] =Tr\,\left[ U\left(
T,0;A^{\prime }\right) \,\,\rho _\varphi \left( 0\right) \,U^{-1}\left(
T,0;A\right) \right]
\end{equation}

To obtain $U\left( T,0;A\right) $ we will use the Heisenberg equation of
motion
\begin{equation}
\label{eq:feq}\ddot \varphi +\left( -i\overrightarrow{\,\nabla }-e\vec A%
\right) ^2\varphi +m^2\varphi =0
\end{equation}

We will choose $\vec A=\left( 0,0,A\left( t\right) \right) \stackrel{}{}$and
\begin{equation}
\label{eq:wy6}\varphi \left( \vec x,t\right) =\sum\limits_{\vec p}\frac 1{%
\sqrt{2\omega _0\left( \vec p\right) V}}\left\{ a_{\vec p}\left( t\right)
+b_{-\vec p}^{+}\left( t\right) \right\} \,e^{i\,\vec p\,\vec x}
\end{equation}

where $a_{\vec p\text{ }}$and $b_{-\vec p}^{+}$ are usual annihilation and
creation operators for particles and antiparticles respectively. In what
follows we will set volume $V=1$. Here $\omega _0\left( \vec p\right) \equiv
$ $\omega \left( \vec p,0\right) $ with
\begin{equation}
\label{eq:wy7}\omega ^2\left( \vec p,t\right) =\vec p^2+\left(
p^3-e\,A\left( t\right) \right) ^2+m^2
\end{equation}

It is well known (see \cite{pvs} ) that the solution of (\ref{eq:feq}) for
creation and annihilation operators has the following form
\begin{equation}
\label{eq:bgl}a_{\vec p}\left( t\right) =\alpha _{\vec p}\left( t\right)
\,a_{\vec p}\left( 0\right) +\beta _{\vec p}^{*}\left( t\right) \,b_{-\vec p%
}^{+}\left( 0\right) \,\,;\,\,b_{-\vec p}^{+}\left( t\right) =\beta _{\vec p%
}\left( t\right) \,a_{\vec p}\left( 0\right) +\alpha _{\vec p}^{*}\left(
t\right) \,b_{-\vec p}^{+}\left( 0\right)
\end{equation}

where the Bogoliubov coefficients $\alpha _{\vec p}\left( t\right) $ and $%
\beta _{\vec p}$ are expressed via auxiliary function $\eta _{\vec p}\left(
t\right) $ by following relations
\begin{equation}
\label{eq:wy8}
\begin{array}{c}
\alpha _{
\vec p}\left( t\right) =\frac 1{2\omega _0\left( \vec p\right) }\left[ i\,%
\frac{d\eta _{\vec p}\left( t\right) }{dt}+\omega _0\left( \vec p\right)
\eta _{\vec p}\left( t\right) \right] \\ \beta _{\vec p}^{*}\left( t\right) =%
\frac 1{2\omega _0\left( \vec p\right) }\left[ i\,\frac{d\eta _{\vec p%
}^{*}\left( t\right) }{dt}+\omega _0\left( \vec p\right) \eta _{\vec p%
}^{*}\left( t\right) \right]
\end{array}
\end{equation}

with the $\eta _{\vec p}\left( t\right) $ satisfying to equation
\begin{equation}
\label{eq:wy9}\frac{d^2}{dt^2}\eta _{\vec p}\left( t\right) +\omega ^2\left(
\vec p,t\right) \,\eta _{\vec p}\left( t\right) =0;\,\,\eta _{\vec p}\left(
t\right) =e^{-i\,\omega _0\left( \vec p\right) \,t}\,\text{ at }t\rightarrow
0\,\,
\end{equation}

Using last equations it is easy to show that $\alpha _{\vec p}\left(
t\right) $ and $\beta _{\vec p}\left( t\right) $ obey to a system of
ordinary first order differential equations
\begin{equation}
\label{eq:wy10}\dot \alpha _{\vec p}\left( t\right) =-i\,h_{\vec p%
}(t)\,\alpha _{\vec p}\left( t\right) -i\,g_{\vec p}(t)\,\beta _{\vec p%
}\left( t\right) \,\,;\,\,\,\dot \beta _{\vec p}(t)=i\,g_{\vec p}(t)\,\alpha
_{\vec p}\left( t\right) +i\,h_{\vec p}(t)\,\beta _{\vec p}\left( t\right)
\end{equation}

with
\begin{equation}
\label{eq:ss1}\,h_{\vec p}(t)=\frac 1{2\omega _0\left( \vec p\right) }\left(
\omega ^2\left( \vec p,t\right) +\omega _0^2\left( \vec p\right) \right)
\,\,\,;\,\,\,g_{\vec p}(t)\,=\frac 1{2\omega _0\left( \vec p\right) }\left(
\omega ^2\left( \vec p,t\right) -\omega _0^2\left( \vec p\right) \right)
\end{equation}

Equations (\ref{eq:wy8}-\ref{eq:ss1}) enable one to consider the particle
creation in a homogeneous electric field $E(t)$ having an arbitrary time
dependence. For example the famous Schwinger results \cite{pair}, describing
pair creation in a constant electric field, can be found using this
equations \cite{pvs}. Note that the number of created particles in $\vec p$
th mode is given by
\begin{equation}
\label{eq:ss3}n_{\vec p}=\left| \beta _{\vec p}\right| ^2
\end{equation}

Now we have enough formulae to find $U(t,0)$ using relations
\begin{equation}
\label{eq:ss2}
\begin{array}{c}
\,a_{
\vec p}\left( t\right) =U^{+}a_{\vec p}U=\alpha _{\vec p}\,a_{\vec p}+\beta
_{\vec p}^{*}\,b_{-\vec p}^{+}\,\,; \\ b_{-\vec p}^{+}\left( t\right)
=U^{+}\,b_{-\vec p}^{+}\,U=\beta _{\vec p}\,a_{\vec p}\,+\alpha _{\vec p%
}^{*}\,b_{-\vec p}^{+}
\end{array}
\end{equation}

here we take for brevity $a_{\vec p}\equiv a_{\vec p}(0)\,;\,b_{-\vec p%
}^{+}\,\,\equiv b_{-\vec p}^{+}\left( 0\right) $.

We will skip the details of calculations and express our result in the
following form $U=\prod_{\vec p}U_{\vec p}$ where $U_{\vec p}$ is (we drop
the mode label)
\begin{equation}
\label{eq:u}U=S(r,\phi )\,R(\theta )
\end{equation}

where
\begin{equation}
\label{eq:ss4}S(r,\phi )=\exp \left[ r\left( e^{2i\phi
}\,a^{+}b^{+}-e^{-2i\phi }\,a\,b\right) \right] \,;\,R(\theta )=\exp \left[
i\,\theta \,\left( a^{+}a+b^{+}b+1\right) \right]
\end{equation}

$S$ and $R$ are called two-mode squeeze and rotation operators respectively.
The parameters $r\,,\phi ,\theta $ are determined from the equations
\begin{equation}
\label{eq:ss5}\alpha =e^{i\,\theta }\,\cosh r\,;\,\beta =e^{i\,\theta
\,-\,\,2i\,\phi }\sinh r
\end{equation}

This expression for $U\left( t,0\right) $ may be useful in many situations.
We may ,for example, rather easily describe time evolution of density matrix
$\rho \left( t\right) $ from arbitrary initial $\rho \left( 0\right) $ ,
more interesting initial states are : vacuum state, thermal equilibrium and
coherent states. Recently the time evolution of density matrix at finite
temperature was considered in \cite{swed} using the functional Schrodinger
representation. In this paper we will deal only with vacuum initial state.
Applying (\ref{eq:u1}) and (\ref{eq:u}) we find that the influence
functional with vacuum as an initial state is given by
\begin{equation}
\label{eq:ef1}\Phi \left[ A^{\prime },A\right] =\prod_{\vec p}\frac 1{\alpha
_{\vec p}^{^{\prime }*}\,\alpha _{\vec p}-\beta _{\vec p}^{^{\prime
}*}\,\beta _{\vec p}}
\end{equation}

\section{Semiclassical CTP effective action and Langevin equation}\label{f4}

We may now, using (\ref{eq:eff}) and (\ref{eq:ef1}), obtain for $\Gamma _{CTP%
\text{ }}$:
\begin{equation}
\label{eq:z1}
\begin{array}{c}
\Gamma _{CTP}\left[ A^{\prime },A\right] =S\left[ A^{\prime }\right]
-S\left[ A\right] +S_{IF}\left[ A^{\prime },A\right] \\
\text{where }\,\,\,\,\,\,\,\,\,\,\,\,\,\,\,S_{IF}\left[ A^{\prime },A\right]
=i\,\sum_{\vec p}\ln \left[ \alpha _{\vec p}^{^{\prime }*}\,\alpha _{\vec p%
}-\beta _{\vec p}^{^{\prime }*}\,\beta _{\vec p}\right]
\end{array}
\end{equation}

It is useful to introduce new variables as
\begin{equation}
\label{eq:ss6}\Xi =\frac 12\left( A^{\prime }+A\right) ;\,\,\,\,\,\,\Delta
=A^{\prime }-A
\end{equation}

and define \cite{q2}
\begin{equation}
\label{eq:z2}C_n(t_1,...,t_n;\Xi _{t_1,0},...,\Xi _{t_n\,,0}]\equiv \frac 1{%
i^{n-1}}\frac{\delta ^n\,S_{IF}\left[ A^{\prime },A\right] }{\delta \Delta
\left( t_1\right) \ldots \delta \Delta \left( t_n\right) } \bigg|_{\Delta=0}
\end{equation}

The notation of $\,\,\,\,C_1(t_1;\Xi _{t_1,0}]$ means $C_{1\text{ }}$is a
function of $t_1$ and also a functional of $\Xi $ between endpoints $t_1$
and $0$. By virtue of (\ref{eq:efff}) the $C_n$'s are real quantities. As
shown in \cite{q2} the $C_n\,$'s can be interpreted as cumulants of
stochastic force. We will ignore all cumulants with $n>2$, because the
cumulants are of order $e^n$ .This will mean that we are making a Gaussian
approximation to the noise. The $\Gamma _{CTP}$ is now
\begin{equation}
\label{eq:d}e^{i\,\Gamma _{CTP}\left[ A^{\prime },A\right] }=e^{i\,S\left[
A^{\prime }\right] \,-\,i\,S\left[ A\right] \,+\,\,i\int\limits_0^\infty
d\tau _1\Delta \left( \tau _1\right) \,C_1(\tau _1;\Xi _{\tau _1,0}]}\,\,e^{-%
\frac 12\int\limits_0^\infty d\tau _1\int\limits_0^\infty d\tau _2\Delta
\left( \tau _1\right) \Delta \left( \tau _2\right) C_2(\tau _1\,,\tau _2;\Xi
_{\tau _1,0}\,,\Xi _{\tau _2,0}]}
\end{equation}

The last term in (\ref{eq:d}) may be recognized as a characteristic
functional of a stochastic gaussian process and we rewrite it in a form
\begin{equation}
\label{eq:ss7}
\begin{array}{c}
\exp \left\{ -
\frac 12\int\limits_0^\infty d\tau _1\int\limits_0^\infty d\tau _2\Delta
\left( \tau _1\right) \Delta \left( \tau _2\right) C_2\left( \tau _1\,,\tau
_2;\Xi _{\tau _1,0}\,,\Xi _{\tau _2,0}\right] \right\} = \\ =\int D\xi
\,P\left[ \xi ,\Xi \right] \,\exp \left\{ i\int\limits_0^\infty d\tau \Delta
\left( \tau \right) \xi \left( \tau \right) \right\}
\end{array}
\end{equation}

where
\begin{equation}
\label{eq:ss8}\,P\left[ \xi ,\Xi \right] =\exp \left\{ -\frac 12%
\int\limits_0^\infty d\tau _1\int\limits_0^\infty d\tau _2\xi \left( \tau
_1\right) C_2^{-1}\left( \tau _1\,,\tau _2;\Xi _{\tau _1,0}\,,\Xi _{\tau
_2,0}\right] \xi \left( \tau _2\right) \right\}
\end{equation}

is the distribution of colour noise $\xi $ .We can use now $\Gamma
_{CTP}\left[ A^{\prime },A\right] $ to obtain the semiclassical Langevin
equations of motion
\begin{equation}
\label{eq:st}\ddot A\left( t\right) =\,\,C_1(t;\Xi _{t,0}]+\xi \left(
t\right)
\end{equation}

In our case we have from (\ref{eq:z1}) and (\ref{eq:z2})
\begin{equation}
\label{eq:ss9}
\begin{array}{c}
\,\,C_1(t;\Xi _{t,0}]=e\int
\frac{d^3\vec p}{\left( 2\pi \right) ^3\omega _0\left( \vec p\right) }\left(
p^3-eA\left( t\right) \right) \left| \alpha _{\vec p}\left( t\right) +\beta
_{\vec p}\left( t\right) \right| ^2\,\,\,\,\,\,\,\text{and} \\ C_2\left(
\tau _1\,,\tau _2;\Xi _{\tau _1,0}\,,\Xi _{\tau _2,0}\right] =\left\langle
\xi \left( \tau _1\right) \xi \left( \tau _2\right) \right\rangle = \\
=\frac{e^2}2\sum_{\vec p}\frac{\left( p^3-eA\left( \tau _1\right) \right)
\left( p^3-eA\left( \tau _2\right) \right) }{\omega _0^2\left( \vec p\right)
}\left\{ \eta _{\vec p}^2\left( \tau _1\right) \eta _{\vec p}^{*2}\left(
\tau _2\right) +\eta _{\vec p}^{*2}\left( \tau _1\right) \eta _{\vec p%
}^2\left( \tau _2\right) \right\}
\end{array}
\end{equation}

The $C_1(t;\Xi _{t,0}]$ is divergent and we must renormalize (\ref{eq:st})
.As in \cite{klu} we will use adiabatic regularization method (or n-wave
regularization of Zel'dovich )\cite{Z,bi}. The advantage of using adiabatic
regularization lies in the fact that we can remove divergences before
summing over modes. Potentially divergent terms does not appear and all
integrals are finite. This is especially useful in numerical computations.
Representing $\eta _{\vec p}\left( t\right) $ in the form
\begin{equation}
\label{eq:ss10}\eta _{\vec p}\left( t\right) =\sqrt{\frac{\omega _0\left(
\vec p\right) }{\Omega _{\vec p}\left( t\right) }}\exp \left\{
-i\,\int\limits_0^t\,\Omega _{\vec p}\left( t^{\prime }\right) dt^{\prime
}\right\}
\end{equation}

we see ,that to perform renormalization, we need only in few first terms in
adiabatic expansion of $\Omega :$%
\begin{equation}
\label{eq:ii1}\frac 1\Omega =\frac 1\omega +\left[ \frac{\ddot \omega }{%
4\omega ^2}-\frac 38\frac{\dot \omega ^2}{\omega ^5}\right] +\ldots
\end{equation}

We rewrite now (\ref{eq:st}) in the form
\begin{equation}
\label{eq:ii2}
\begin{array}{c}
\ddot A\left( t\right) =e\int \frac{d^3\vec p}{\left( 2\pi \right) ^3}\left(
p^3-eA\left( t\right) \right) \left[ \frac 1{\Omega _{\vec p}\left( t\right)
}-\frac 1{\tilde \Omega \left( t\right) }+\frac 1{\tilde \Omega \left(
t\right) }\right] \,\,+\xi \left( t\right) = \\ =e\int \frac{d^3\vec p}{%
\left( 2\pi \right) ^3}\left( p^3-eA\left( t\right) \right) \left[ \frac 1{%
\Omega _{\vec p}\left( t\right) }-\frac 1{\tilde \Omega \left( t\right) }%
\right] \,-\ddot A\left( t\right) \frac{e^2}{48\pi ^2}\ln \frac{\Lambda ^2}{%
m^2}\,+\xi \left( t\right)
\end{array}
\end{equation}

where
$$
\frac 1{\tilde \Omega \left( t\right) }\equiv \frac 1\omega +\left[ \frac{%
\ddot \omega }{4\omega ^2}-\frac 38\frac{\dot \omega ^2}{\omega ^5}\right]
\,\,\text{and }\Lambda \text{ is a cutoff in the transverse momentum
integration.}
$$

Remembering now that in scalar QED the usual charge renormalization factor Z$%
_3$ is equal to
\begin{equation}
\label{eq:ii3}Z_3=1-\frac{e^2}{48\pi ^2}\ln \frac{\Lambda ^2}{m^2}\,\,;\,%
\text{and that \thinspace \thinspace }e_r=Z_3^{\frac 12}\,e\,\,\,;\,\,%
\,A_r=Z_3^{-\frac 12}\,A
\end{equation}

we obtain
\begin{equation}
\label{eq:fini}
\begin{array}{c}
\ddot A_r\left( t\right) =e_r\Psi \left[ e_rA_r\right] +\xi _r\,\, \\ \text{%
with }\Psi \left[ e_rA_r\right] \equiv e_r\int \frac{d^3\vec p}{\left( 2\pi
\right) ^3}\left( p^3-e_rA_r\left( t\right) \right) \left[ \frac 1{\Omega _{%
\vec p}\left( t\right) }-\frac 1{\tilde \Omega \left( t\right) }\right]
;\,\,\xi _r=Z_3^{\frac 12}\xi
\end{array}
\end{equation}

In $C_2$ we must also change $e$ to $\,e_r.$ So we obtain the finite
renormalized Langevin equation (\ref{eq:fini}) that describe the process of
pair production in a spatially homogeneous electric field and the
backreaction from this pairs on time evolution of the electric field. The
solution of the Langevin equation is beyond the scope of the present
paper.We plan to consider this solution in future. Notice that without noise
term Eq.(\ref{eq:fini}) is equal to the semiclassical Maxwell equation,
obtained in \cite{klu}.

\section{Decoherence in scalar QED}\label{f5}

In this section we will show ,using the results of previous sections, how
the programme of decoherence \cite{gh,zu} can be applied in the context of
quantum electrodynamics in some detail. We then analyze an example where
macroscopic electromagnetic fields are ``measured'' through interaction with
charges and thereby rendered classical. This example was discussed recently
by Kiefer \cite{kief} using different point of view.

In the consistent or decoherent histories formulation of quantum mechanics
\cite{quant1,quant2,quant3}the complete description of a coupled $\psi
,\sigma $ system is given in terms of fine-grained histories $\psi \left(
t\right) ,\sigma \left( t\right) .$ Let us take as a coarse-graining
procedure of summing over the $\sigma $ field. In other words the $\sigma $
field play in our case the role of environment. Then the interference
effects between coarse-grained histories are measured by the decoherence
functional $D\left[ \psi ,\psi ^{\prime }\right] $. It was shown in \cite
{q1,hu,sinha} that the decoherence functional, which is the fundamental
object of the decoherent histories formulation, is connected with CTP\
effective action by following relation
\begin{equation}
\label{eq:keq}D\left[ \psi ,\psi ^{\prime }\right] =e^{i\,\Gamma
_{CTP}\,\,\left[ \psi \,\,,\,\psi ^{\prime }\right] }
\end{equation}

The coarse-grained history $\psi \left( t\right) $ can be described
classically if and only if the decoherence functional is approximately
diagonal , that is, $D\left[ \psi ,\psi ^{\prime }\right] \simeq 0$ whenever
$\psi \neq \psi ^{\prime }.$ Now after this very brief discussion of the
some aspects of decoherence,we will proceed to discuss an example where
macroscopic field strengths decohere through their interaction with
charges.We wish to consider, as an example, a macroscopic superposition of
two electric fields, one pointing upwards, and the other pointing downwards.
In the case of scalar QED we have
\begin{equation}
\label{eq:kk1}D\left[ A,A^{\prime }\right] =e^{i\,\Gamma _{CTP}\left[
A^{\prime },A\right] }\sim \,\,e^{-\frac 12\int\limits_0^\infty d\tau
_1\int\limits_0^\infty d\tau _2\Delta \left( \tau _1\right) \Delta \left(
\tau _2\right) C_2(\tau _1\,,\tau _2;\Xi _{\tau _1,0}\,,\Xi _{\tau _2,0}]}
\end{equation}

Here we have omitted unessential phase factor. Now after an easy calculation
we obtain
\begin{equation}
\label{eq:kk2}D\left[ A,A^{\prime }\right] \sim \exp \left\{ -\frac{%
V\,e^2\,E^2}{256\pi m}\right\}
\end{equation}

This result resembles the result obtained by Kiefer. Note that the
interaction with the charge states leads to an exponential suppression
factor of the corresponding interference terms for the field; in the
infrared limit of $V\to \infty $ one finds exact decoherence. In realistic
cases, however, a finite coherence width remains, so one can in principle
subject these results to experimental confirmation. For an electric field of
$E\approx 10^7$ Volts per centimeter, for example, one finds that
interference effects are observable on length scales $L\leq 10^{-4}$
centimeters \cite{kief}.

Thus the programme of decoherence \cite{quant2} may successfully be applied
in the context of quantum field theory using the concepts and methods of
nonequilibrium statistical field theory.

{\bf Acknowledgments}

I would like to thank V.N.Baier for his interest in this work and valuable
comments on the text of the manuscript.

\end{document}